\documentclass[a4paper,fleqn,usenatbib]{mnras}

\usepackage[T1]{fontenc}
\usepackage{ae,aecompl}
\usepackage[dvipdfm]{graphicx}

\usepackage{graphicx}	% Including figure files
\usepackage{amsmath}	% Advanced maths commands
\usepackage{amssymb}	% Extra maths symbols

\usepackage{grffile}
\usepackage{longtable}
\usepackage{rotating}
\usepackage{lscape}

\title[The VMC survey - XXIII. Model fitting of SMC classical
Cepheids]{The VMC survey - XXIII. Model fitting of light and radial
  velocity curves of Small Magellanic Cloud classical Cepheids\thanks{Based on observations collected at the European Organisation for Astronomical
Research in the Southern Hemisphere under ESO programme(s) 179.B-2003.}}

\author[Marconi et al.]{M. Marconi,$^{1}$\thanks{E-mail:  marcella.marconi@oacn.inaf.it}
R. Molinaro,$^{1}$, V. Ripepi$^{1}$, M.-R. L. Cioni$^{2,3}$, G. Clementini$^{4}$
\newauthor{M. I. Moretti$^{1}$, F. Ragosta$^{1}$, R. de Grijs$^{5,6}$,
  M.~A.~T.~Groenewegen$^{7}$, V. D. Ivanov$^{8,9}$}\\
$^{1}$INAF-Osservatorio Astronomico di Capodimonte, Salita
  Moiariello, 16, I-80131, Napoli, Italy\\
$^{2}$Leibnitz-Institut f\"ur Astrophysik Potsdam, An der Sternwarte 16, D-14482 Potsdam, Germany\\
$^{3}$University of Hertfordshire, Physics Astronomy and Mathematics,
College Lane, Hatfield AL10 9AB, UK\\
$^{4}$INAF - Osservatorio Astronomico di Bologna, Via Ranzani 1, I-40127 Bologna, Italy\\
$^{5}$Kavli Institute for Astronomy \& Astrophysics and Department of
Astronomy, Peking University, Yi He Yuan Lu 5,\\ Hai Dian District, Beijing 100871, China\\
$^{6}$International Space Science Institute--Beijing, 1 Nanertiao, Zhongguancun, Hai Dian District, Beijing 100190, China\\
$^{7}$Koninklijke Sterrenwacht van Belgi\"e, Ringlaan 3, B-1180 Brussels, Belgium\\
$^{8}$European Southern Observatory, Ave. Alonso de C\'ordova 3107, Vitacura, Santiago, Chile\\
$^{9}$European Southern Observatory, Karl-Schwarzschild-Strasse 2, D-85748 Garching bei M\"unchen, Germany}

\date{Accepted XXX. Received YYY; in original form ZZZ}

\pubyear{2016}

\begin{document}
\label{firstpage}
\pagerange{\pageref{firstpage}--\pageref{lastpage}}
\maketitle

\begin{abstract}
We present the results of the ${\chi}^2$ minimization model fitting technique applied to
optical and near-infrared photometric and radial velocity data for a
sample of 9 fundamental and 3 first overtone classical Cepheids in the Small
Magellanic Cloud (SMC). The near-infrared photometry ({\it JK} filters) was
obtained by the  European Southern Observatory (ESO) public survey ``VISTA near-infrared $Y, J, K_\mathrm{s}$ 
survey of the Magellanic Clouds system'' (VMC).
For each pulsator isoperiodic model sequences have been computed by
adopting a nonlinear convective hydrodynamical code in order to
reproduce the multi-filter light and (when available) radial velocity
curve amplitudes and morphological details. 
The inferred individual distances provide an intrinsic mean value for the SMC distance modulus of 19.01 mag and a standard deviation of 0.08 mag, in agreement with the literature.
Moreover the instrinsic masses and luminosities of the best fitting model
show that all these pulsators are brighter than the canonical
evolutionary Mass-Luminosity relation (MLR), suggesting a significant
efficiency of core overshooting and/or mass loss. Assuming that
the inferred deviation from the canonical MLR is only due to
mass loss, we derive the expected distribution of percentage mass loss as a
function of both the pulsation period and of the canonical stellar mass. 
Finally, a good agreement is found between the predicted mean
radii and current Period-Radius (PR) relations in the SMC available in
the literature.
The results of this investigation support the predictive capabilities
of the adopted theoretical scenario and pave the way to the 
application to other extensive databases at various chemical
compositions, including the VMC  Large
Magellanic Cloud pulsators and Galactic Cepheids with Gaia parallaxes. 
\end{abstract}


%%%%%%%%%%%%%%%%%%%%%%%%%%%%%%%%%%%%%%%%%%%%%%%%%%

%%%%%%%%%%%%%%%%% BODY OF PAPER %%%%%%%%%%%%%%%%%%
\section{Introduction}

Classical Cepheids (CCs) are
widely adopted as primary distance indicators to calibrate the
extragalactic distance scale. Indeed, several secondary
distance indicators, which
are able to directly constrain the Hubble
constant, rely on Cepheid Period-Luminosity (PL) and
Period-Luminosity-Colour (PLC) relations \citep[see e.g.][and
references therein]{fried01,saha01,riess16}.
From the evolutionary point of view, Cepheids are intermediate-mass
stars during the central helium burning phase \citep[see e.g.][and
references therein]{bono00a,valle09}, crossing the instability strip
as they move bluewards in the Hertzprung-Russell (HR)
diagram ({\it blue loop} excursion) at constant luminosity for each
given mass. Thus, stellar evolution predicts  a Mass-Luminosity
relation (MLR) for classical Cepheids.
This MLR, combined with the period-density relation
and the Stefan Boltzmann law, produces a PLC relation, holding for each
individual Cepheid \citep[see e.g.][and references
therein]{bono99,marconi05,marconi09}.
Then,  projecting the PLC onto the PL plane gives the PL relation
\citep[see e.g.][]{madore91,bono99,marconi09}. On this basis, the investigation of Cepheid pulsation properties is also crucial for providing independent constraints on the MLR.
Several authors have discussed the uncertainties related to this MLR as due to core overshooting and/or mass loss
\citep[see e.g.][and references
therein]{chiosi93,was97,bono99,kw06,caputo05,marconi13b,musella16}, often
related to the so called {\it mass discrepancy} problem \citep[see
e.g.][and references therein, for details]{marconi13b}, first
suggested by \citet{stobie69} and \citet{christy70}. 
These authors noticed that the Cepheid {\it evolutionary} masses based on  the application of theoretical
isochrones to observations were larger than the {\it pulsational}
ones, e.g. based on period–mass–radius relations \citep{Fricke71,bono01}.
One of the methods that has been recently adopted by our team to address
the mass discrepancy problem is the so called model fitting of
multi-filter light, radial velocity and radius curves \citep[see
e.g.][and references therein]{kw06,natale08,marconi13a,marconi13b},
through the direct comparison of the observed  and predicted variations along a
pulsation cycle, the latter based on nonlinear convective pulsation models
\citep[see][for details]{bcm00,bcm02,marconi13b}. 
In particular, beyond the application to the prototype $\delta$ Cephei
\citep[see][]{natale08}, three papers have been devoted by our team to the model
fitting of variations for Cepheids in the Large Magellanic Clooud:

\begin{itemize}

\item In \citet{bcm02} we fitted the V, I band light curves of two LMC bump
Cepheids from the OGLE database \citep{Udalski99}, with the bump
(secondary maximum) along either the decreasing (OGLE
194103, shorter-period Cepheid) or the rising (OGLE 56087,
longer-period Cepheid)
branch of the light curve.  The adopted nonlinear convective pulsation
models reproduced the luminosity variation over the entire pulsation
cycle if the adopted stellar mass for both Cepheids  was roughly 15\%
smaller than predicted by evolutionary models neglecting mass loss and
convective core overshooting. Moreover, the model fitting procedure
provided a distance modulus to the LMC of 18.53$\pm$0.05 mag. This
value was in excellent agreement with the results based on the model
fitting technique applied by other teams \citep[see e.g.][and
references therein]{kw06} to LMC Cepheids from the MACHO \citep{alcock99} database, as well as with the distance
obtained from application of the model fitting technique to RR Lyrae
\citep{mc05} or $\delta$ Scuti \citep{mcna07} pulsators. 

\item In \citet{marconi13a} we fitted the multifilter (U, B, V, I and
  K) light and radial velocity curves of five Cepheids in
  NGC 1866, an LMC young massive cluster. Again, the inferred
  stellar parameters corresponded to a MLR slightly brighter
  than the canonical one, as an effect of mild overshooting and/or
  mass loss, and to individual distances consistent within the
  uncertainties. The resulting mean distance modulus (18.56 $\pm$ 0.03
  (stat) $\pm$ 0.1 (syst) mag), was found to be in agreement with the
  literature, as well as with the previous model fitting applications.

\item In \citet{marconi13b} we fitted the light and radial velocity
  curves of the LMC Cepheid OGLE-LMC-CEP-0227 belonging to a detached
  double-lined eclipsing binary system, finding, for the best fitting model,  a pulsation mass, a
  mean effective temperature, a luminosity amplitude and a mean radius
  in agreement with the empirical estimates. The inferred MLR was again more in agreement with
  evolutionary models including a moderate amount of overshooting
  and/or mass loss \citep[][]{cs11,prada12}, the best fitting
  chemical composition was more metal-poor than typical LMC
  Cepheids (Z = 0.004 versus 0.008) and slightly helium enhanced (Y =
  0.27 versus 0.25) and the inferred true distance modulus of the
  LMC (18.50 $\pm$ 0.02 $\pm$ 0.10 (syst) mag), was found to be  in excellent agreement
  with similar estimates from independent methods in the literature.

\end{itemize}

In this paper we extend previous analyses to a sample of
9 fundamental (F) and 3 first overtone (FO) CCs in the Small Magellanic Cloud (SMC), that are also targets of
the ``VISTA near-infrared YJKs survey of the Magellanic Clouds System''
\citep[VMC, P.I.: M.-R. L. Cioni; see][]{cioni11}.

The organization of the paper is the following. In Section 1 we
discuss the sample selection with a brief outline of the VMC survey.
In Section 2 we recall the main steps of the adopted model fitting
technique and in Section 3 its application to the selected SMC
Cepheids.
Section 4 deals with the implications of the model fitting results for
what concerns the MLR,  the Period-Radius (PR), the
PL and the Wesenheit relations.
The final section includes the summary and some  perspectives.

\section{Selection of the sample}\label{sample}

The selected sample of CCs includes 9 F and 3 FO pulsators with optical photometry from the OGLE III database
\citep{so10} and NIR photometry from the VMC survey
\citep{cioni11,ripe16}. For  Harvard Variable (HV) stars we used
also V, I, J, K data from \citet{storm04}. We trasformed the J
  and K band data by these authors from the California Institute of Technology (CIT) photometric system to
  the VISTA system. To this aim we used equations by \citet{car01} to
  pass from CIT to 2MASS system and then we trasformed 2MASS data to VISTA by using the equations provided by the Cambridge Astronomy Survey Unit (CASU)\footnote{http://www.astro.caltech.edu/~jmc/2mass/v3/transformations/}, \citep[see also][]{ripe16}.

VMC  covers the entire Magellanic
system with deep NIR ($Y, J, K_\mathrm{s}$ filters) VIRCAM  (VISTA InfraRed Camera; Dalton
et al. 2006) photometry on the ESO/VISTA telescope (Emerson et al.
2006). This survey aims at deriving the Star Formation History
and its spatial variation and  an accurate 3D map of
the  Magellanic system by using pulsating stars as distance indicators
and tracers of stellar populations \citep[see e.g.][and references
therein]{ripe12a,ripe12b,ripe14,ripe15,more14,mura15}.
In particular the VMC SMC Cepheids have been presented and discussed by
\citet{ripe16} and \citet{more16}. The average number of phase points
are 5.7,  6.3, and 16.7 in $Y, J, K_\mathrm{s}$, respectively \citep[see][for details]{ripe16}.

The properties of the selected SMC Cepheids are reported in the first six columns of Table~\ref{tab:cepProperties}.  
The first two columns report the identification and the pulsation mode,
whereas the third and fourth column provide the period and epoch
information. The available data are reported in the fifth column (only
photometry or photometry and radial velocity), while the intensity mean magnitude in the V band, derived from the best fitting models, is shown in column six. 

The radial velocity data for the three last variables in
Table~\ref{tab:cepProperties} are taken from \citet{storm04}. They
are derived with the cross correlation method \citep[see][for details]{storm04}.

\begin{figure*}
\includegraphics[width=8.5cm]{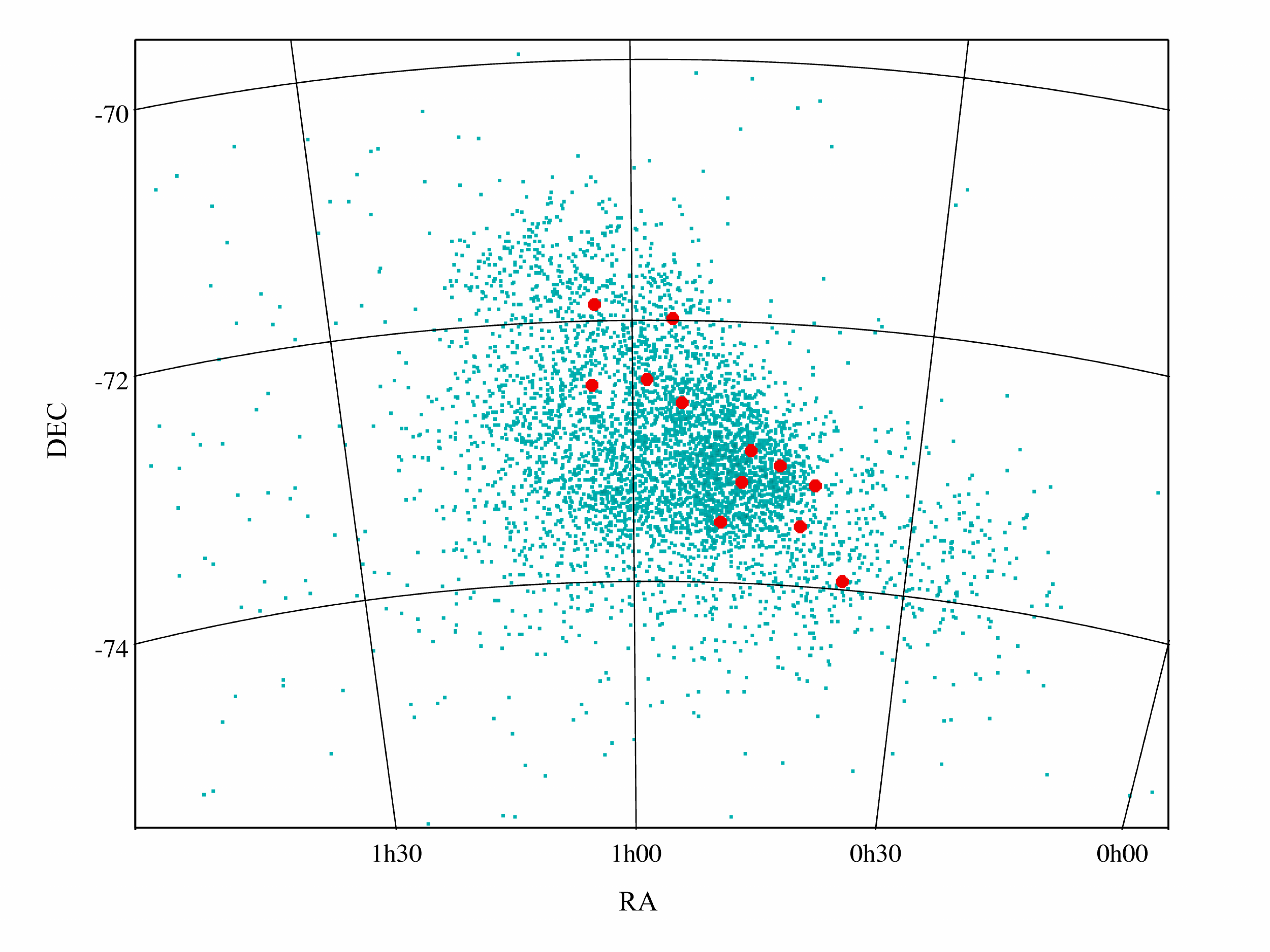}
\caption{The distribution in right ascension and declination of the
  selected CCs. The whole SMC dataset is shown for comparison.}
\label{position}
\end{figure*}

The selected 12 stars were chosen in order to span different periods, amplitudes  and
morphologies of the observed light curves from \citet{ripe16}
database. Even if they do not represent the entire sample, their
properties allow  us to test
the predictive capability of the model fitting technique on a variety
of observed Cepheid properties in the same stellar system. We note
that a statistically significant extension of the selected target
number  would be
extremely time consuming, given the accuracy of the fitting procedure. 
We also note that the selected CCs are well distributed across the SMC
body (see Fig. \ref{position}). 

%===================== TABLES =============================

\begin{landscape}
\begin{table}[h]
\centering
\caption{Properties of the selected SMC Cepheids}
\label{tab:cepProperties}
\begin{tabular}{ccccccccccccccc}
  \hline
ID & mode & P & Epoch & data & $<V>$ & T$_{eff}$ & $\log(L/L_\odot)$ & $M/M_\odot$ & $R/R_\odot$ & $\mu_0  \pm (\delta\mu_0)_{rms}$ & $(\delta\mu_0)_{grid}$ & $p$ & $(\delta p)_{grid}$ & $\gamma \pm (\delta\gamma)_{rms}$ \\ 
 & & (days) & (HJD) & & (mag) & (K) & (dex) & & & (mag) & & & (km/s) & (km/s) \\
  \hline
  \hline
OGLE\,0646 & FO & 1.2781746 & 50621.24375 & phot. & 17.207 & 6425 & 2.78 & 3.6 & 19.9 & 19.01 $\pm$ 0.06 & (-0.02; 0.06) &  &  &\\ 
  OGLE\,1021 & FO & 1.6006633 & 50621.42165 & phot. & 17.171 & 6350 & 2.83 & 3.0& 21.6 & 18.94 $\pm$ 0.03 & (-0.02; 0.02) &  &  &\\ 
  OGLE\,2306 & FO & 1.6063795 & 50621.50367 & phot. & 17.091 & 6450 & 2.86 & 3.0 & 21.6 & 18.93 $\pm$ 0.06 & (-0.02; 0.02) &  &  &\\ 
  OGLE\,1518 & F & 2.8766747 & 50465.34000 & phot. & 16.676 & 6160 & 3.00 & 3.8 & 28.1 & 19.01 $\pm$ 0.12 & (-0.03; 0.01) &  &  &\\ 
  OGLE\,3588 & F & 5.3190652 & 50619.01947 & phot. &15.681  & 5975 & 3.35 & 5.0 & 44.4 & 19.00 $\pm$ 0.10 & (-0.01; 0.03) &  &  &\\ 
  OGLE\,3533 & F & 7.9896097 & 52102.05814 & phot. & 15.163 & 5900 & 3.50 & 4.6 & 54.0 & 18.97 $\pm$ 0.04 & (-0.02; 0.02) &  &  &\\ 
  OGLE\,1165 & F & 10.3090126 & 50612.86364 & phot. & 15.911 & 5700 & 3.66 & 6.0 & 69.4 & 19.23 $\pm$ 0.06 & (-0.01; 0.03) &  &  &\\ 
  HV\,1345 & F & 13.4784 & 47496.0 & phot./v$_{rad}$ & 14.765 & 5325 & 3.60 & 5.0 & 74.2 & 18.97 $\pm$ 0.14 & (-0.02; 0.02) & 1.38 & (-0.05; 0.05) & 105.8 $\pm$ 7.3\\ 
  HV\,1335 & F & 14.3816 & 50610.6 & phot./v$_{rad}$ & 14.809 & 5325 & 3.66 & 5.4 & 79.6 & 19.10 $\pm$ 0.16 & (-0.01; 0.02) & 1.40 & (-0.05; 0.01) & 154.1 $\pm$ 4.9\\ 
  OGLE\,2841 & F & 14.712884 & 50616.27151 & phot. & 14.955 & 5350 & 3.73 & 5.9 & 86.4 & 19.03 $\pm$ 0.08 & (-0.02; 0.01) &  &  &\\ 
  HV\,822 & F & 16.7421 & 47485.9 & phot./v$_{rad}$ & 14.538 & 5320 & 3.77 & 5.9 & 90.6 & 18.98 $\pm$ 0.12 & (-0.03; 0.04) & 1.16 & (-0.03; 0.01) & 99.0 $\pm$ 3.1\\
  OGLE\,2470 & F & 42.7468803 & 52079.01257 & phot. & 13.242 & 5400 & 4.41 & 9.8 & 183.6  & 18.97 $\pm$ 0.11 & (-0.04; 0.05) &  &  &\\ 
   \hline
\end{tabular}
\end{table}
\end{landscape}

\section{The model fitting technique} \label{technique}
The fitting technique, applied to find the best model reproducing the observations, is the same as that adopted in \citet{marconi13a}. The photometric curves of the models have been phased in order to find the V band maximum of light at phase zero. Afterwards, for each modeled photometric band, we estimated the magnitude shift, $\delta M$, which provides the best match with the observed light curves. Specifically, indicating with $m$ and $M_{model}$ the observed apparent magnitude and the model absolute magnitude respectively, we minimized the following $\chi^2$ function:
\begin{equation}\label{eq-chi2_phot}
\chi^2 = \sum_{i=1}^{N_{bands}}\sum_{j=1}^{N_{points}}\left[m^i_j - \left(M^i_{model}(\phi^i_j+\delta\phi^i) + \delta M^i  \right)\right]^2 
\end{equation}
where the two sums are performed over the number of bands,
$N_{bands}$, and the number of measures, $N_{points}$. To estimate the
value of the model magnitude at the same phase of the observations, we
have used a smooth spline interpolation. The fitted parameters are
$\delta{\phi}$, which represents possible small residual ($\sim \pm 0.1$) phase shifts between model and data, and $\delta M^i$, which represents the distance modulus in the $i^{th}$ photometric band. 

For the three Cepheids with radial velocity observations, we
trasformed the modeled pulsational velocity into radial velocity by
using the equation $v^{radial}=(-1/p)v^{model}$, where $p$ is the
projection factor. In this case the $\chi^2$ function assumes the form:
\begin{equation}\label{eq-chi2_vr}
\chi^2=\sum_{j=1}^{N_{points}}\left[v_j - \left(-\frac{1}{p}v^{model}\left(\phi_j +\delta \phi \right) + \gamma\right)\right]^2  
\end{equation}
where $v$ is the observed radial velocity and $\gamma$ is the
barycentric velocity. As in the case of the photometry, we  calculated
the pulsational velocity at the phase of the observed data by using a
smooth spline interpolation. In this case the fitted parameters are
the barycentric velocity and the projection factor as well as the
 phase shift $\delta \phi$ between data and models.

 We notice that the uncertainty affecting the $p$ factor is
still the main source of systematic errors in the various versions of the Baade–
Wesselink method  and its value and possible period dependence are lively
debated in the literature\citep[see e.g.][and references
therein]{molinaro11,marconi13a}. The model fitting  of radial velocity
curves represents an
independent tool to contrain the value of this crucial parameter
\citep[see also][and references therein]{natale08}.

For Cepheids with radial velocity data, we combined the $\chi^2$ functions defined above by normalizing the rms of residuals in each band and radial velocity by their  corresponding pulsation amplitudes, and then we summed in quadrature to obtain a total normalized rms.

\section{Application to the selected SMC Cepheids}\label{results}

Following a similar strategy to the one adopted in
\citet{marconi13a,marconi13b}, we  constructed a
large set of pulsation models by adopting the typical chemical
composition of SMC young stellar population
\citep[Z = 0.004, Y = 0.25; see e.g.][]{luck98,romaniello08}, varying the stellar
parameters mass, luminosity and effective temperature in order to match the
pulsation periods (see Table~\ref{tab:cepProperties}). The output
bolometric light curves were converted into observational bands by
using static model atmospheres by \citet{caste97a,caste97b} and, for
the NIR filters,
subsequent conversion of  Johnson-Cousin magnitudes into
the VISTA photometric system, by using the following equations from \citet{ripe16}:
\begin{eqnarray}
  K^V_S= K^J + 0.007(V-K)^J + 0.03(J-K)^J -0.038 \\ \nonumber
  (V-K_S)^V = 0.993(V-K)^J - 0.03(J-K)^J + 0.038 \\ \nonumber
  (J-K_S)^V = 0.87(J-K)^J -0.01
\end{eqnarray}

\begin{figure*}
\includegraphics[width=15.0cm]{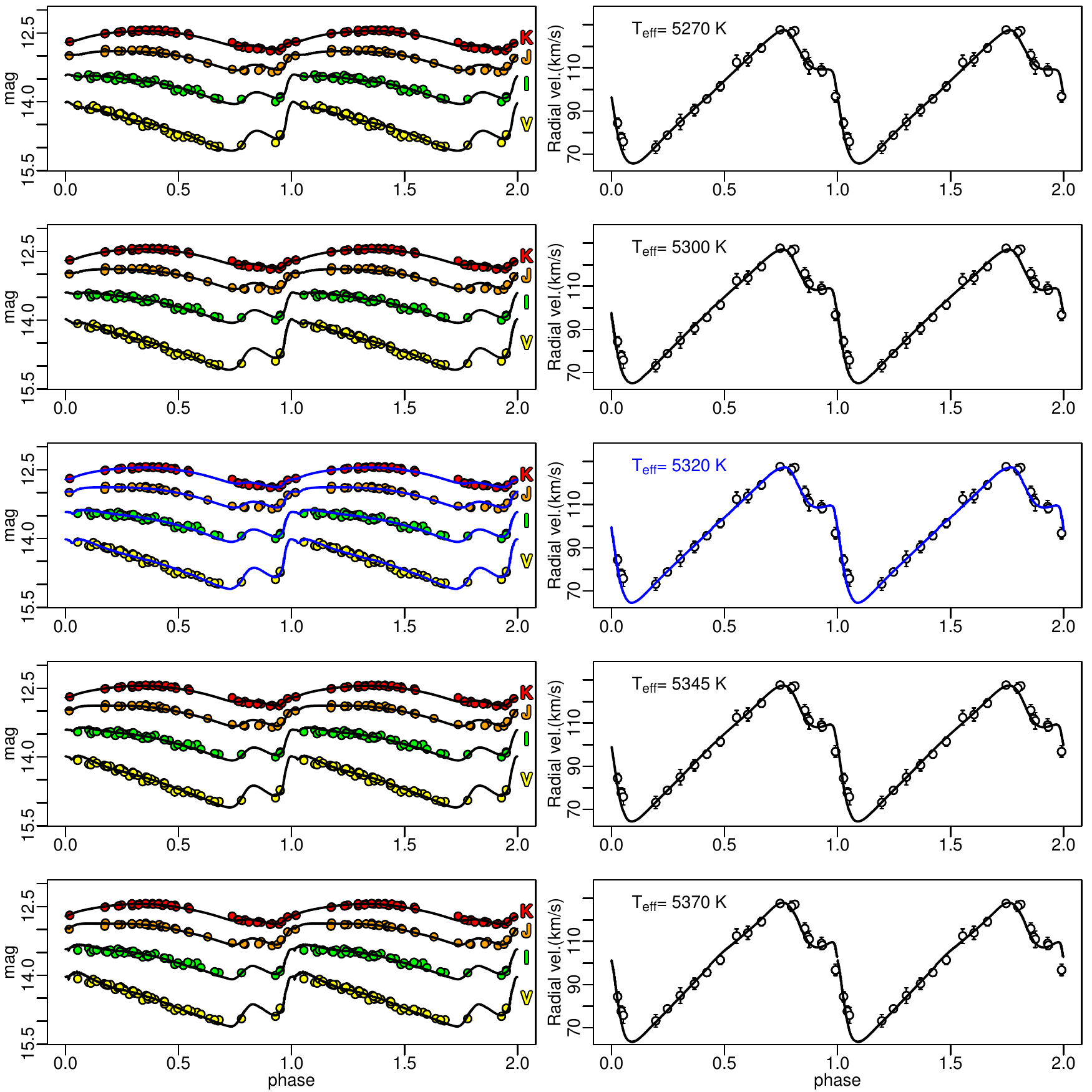}
\caption{Model fitting procedure followed to estimate the best fitting
  effective temperature for the Cepheid HV822. Photometry (left
  panels) and radial velocity (right panels) are plotted for models
  with fixed stellar mass value ($M=7.0M_\odot$) and varied effective
  temperature from $T_{eff}$=5270 K to $T_{eff}$=5370 K. The effective
  temperature of the best fitting model (blue lines) is equal to the
  value that minimizes the $\chi^2$ functions of
  eq.~\ref{eq-chi2_phot} and ~\ref{eq-chi2_vr}. As the mean error
    bar for photometry is always $<0.02$ mag, we do not plot it for clarity reasons.}
\label{fig-modelGrid_T}
\end{figure*}

\begin{figure*}
\includegraphics[width=15.0cm]{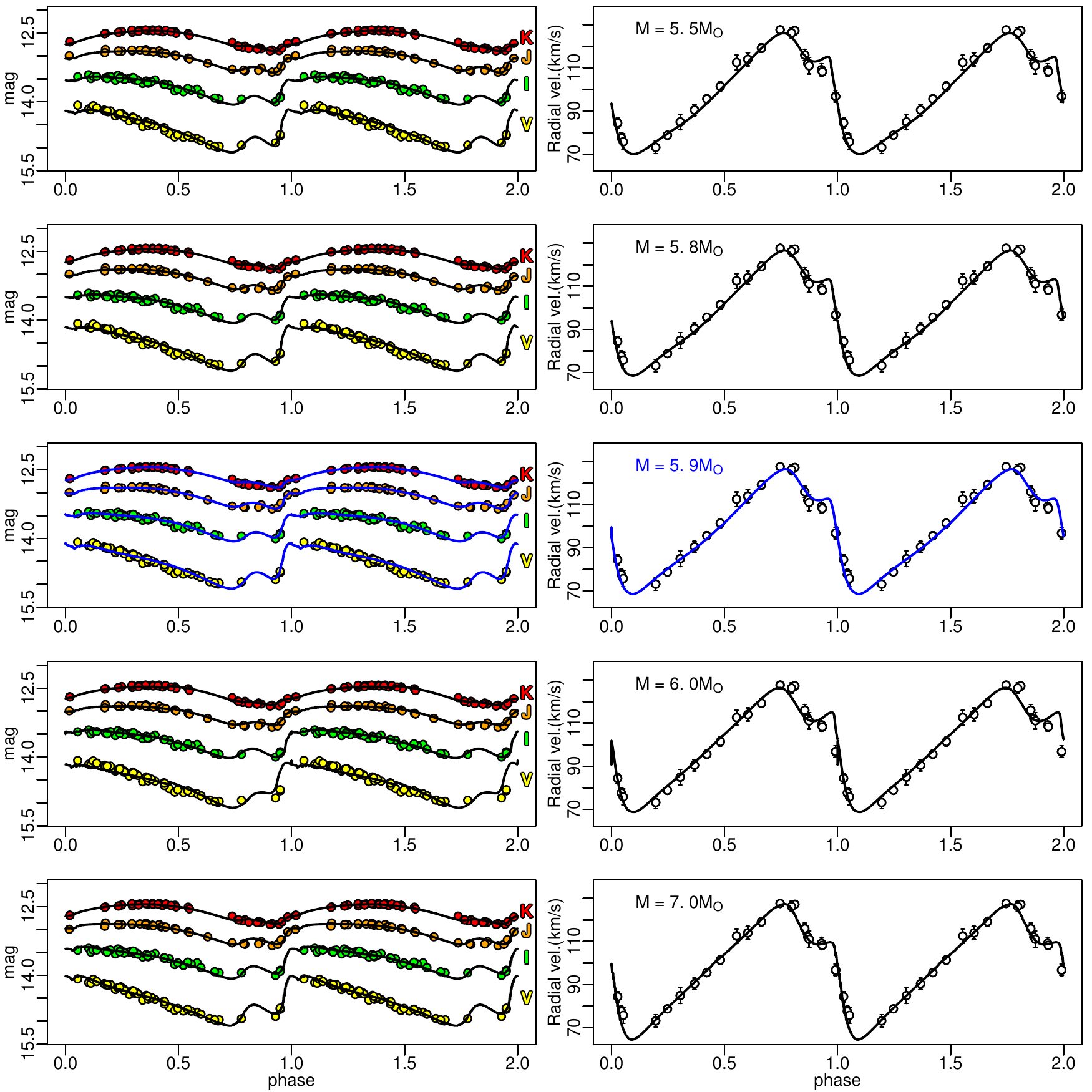}
\caption{Model fitting procedure followed to estimate the best fitting
  mass for the Cepheid HV822. Similarly to Fig.~\ref{fig-modelGrid_T},
  the photometry (left panels) and radial velocity (right panels) are
  plotted for models with effective temperature equal to the best
  fitting value ($T_{eff}=5320 K$) and mass value varying from $M=5.5
  M_\odot$ to $M=7.0 M_\odot$. The mass of the best fitting model
  (blue line in plot) is equal to the value that minimizes the
  $\chi^2$ functions of eq.~\ref{eq-chi2_phot} and
  ~\ref{eq-chi2_vr}. As the mean error
    bar for photometry is always $<0.02$ mag, we do not plot it for clarity reasons.}
\label{fig-modelGrid_M}
\end{figure*}

For each individual pulsator, at the corresponding fixed period we
first assumed a stellar mass and
varied the stellar luminosity and effective temperature.
For example, in the case of variable HV822 (see Table~\ref{tab:cepProperties}), by applying the $\chi^2$ analysis to the models computed at fixed
chemical composition (Z = 0.004,Y = 0.25) and stellar mass
(M = 7.0$M_\odot$), we find that the effective temperature of the
best-fit model is $T_{eff}$ = 5320 $\pm$ 25 K (see Fig.~\ref{fig-modelGrid_T}), corresponding to a
luminosity level of $\log{L/L_{\odot}}=3.84\pm 0.01$.  The uncertainties
on the best-fit intrinsic parameters correspond to the step in 
mass and effective temperature of the different sets of
models.
Then we fixed the effective temperature of the obtained
best fit solution ($T_{eff}=$5320 K) and varied the mass and luminosity with a step of
0.1 in $M/{M_{\odot}}$ and of 0.01 dex in $\log{L/L_{\odot}}$  until the 
best fitting model is obtained by minimization of the combined photometry
and radial velocity ${\chi}^2$ \citep[see discussion in the previous Section
and in][]{marconi13a,marconi13b}.
Figure ~\ref{fig-modelGrid_M} shows the model fitting of variable HV822 by varying the mass and the luminosity at fixed period and effective temperature.
The minimum   ${\chi}^2$  is obtained for $M/{M_{\odot}}=5.9$ and
$\log{L/L_{\odot}}=3.77$.

\begin{figure*}
\includegraphics[trim=0cm 9.5cm 0cm 0cm,width=15.0cm]{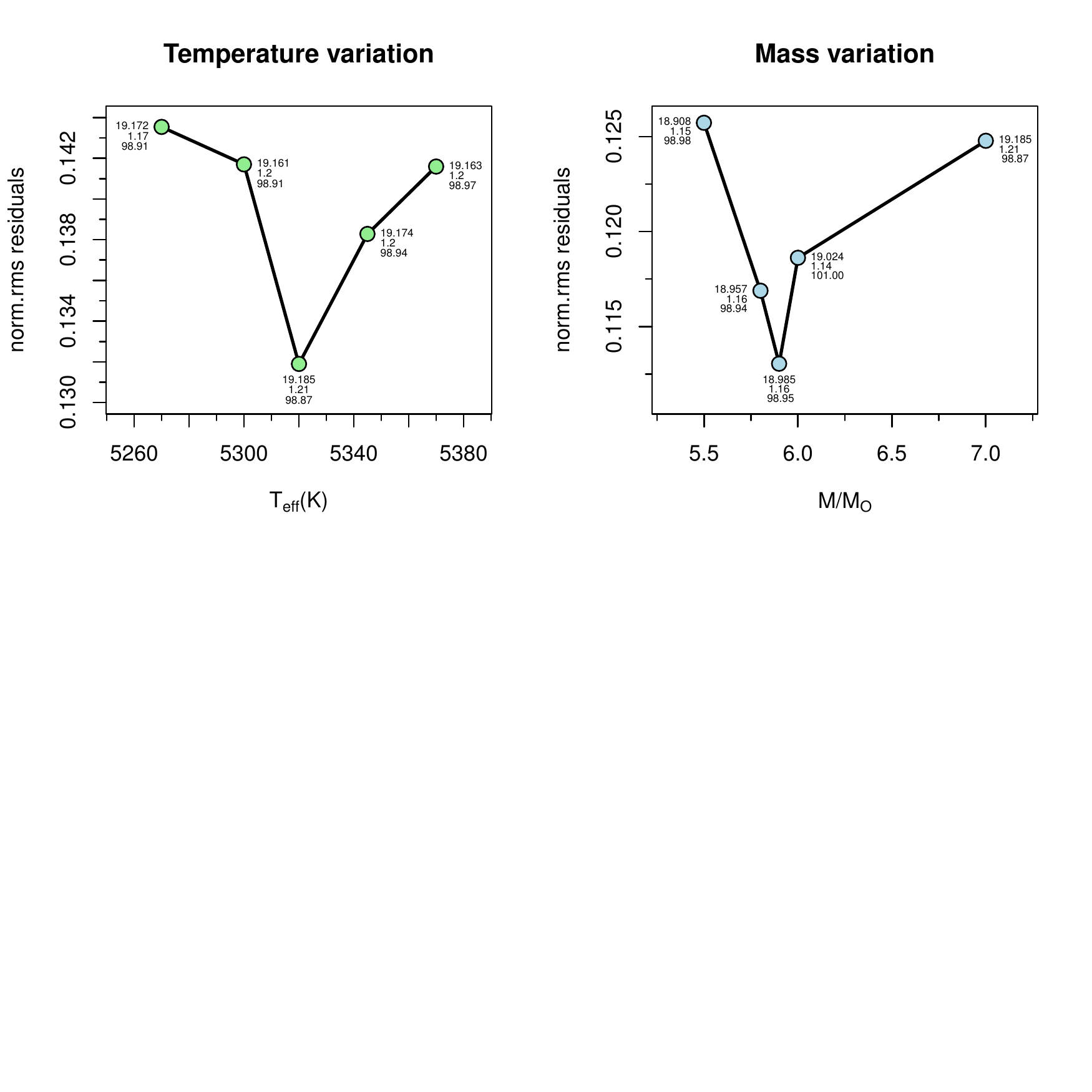}
\caption{The normalized rms of residuals obtained from the fitting
  procedure applied to HV822 are shown asa  function of the model effective temperature (left panel) and mass (right panel). The inferred true distance modulus $\mu_0$, the $p$ factor and the
barycentric velocity $\gamma$ are labelled for each dot.}
\label{fig-normRmsResiduals}
\end{figure*}

Since it is difficult to evaluate by-eye the quality of the fit
  from Fig.~\ref{fig-modelGrid_T} and Fig.~\ref{fig-modelGrid_M}, we
  show in Fig.~\ref{fig-normRmsResiduals} the normalized rms of
  residuals as a function of the model effective temperature (left
  panel) and mass (right panel). The plotted rms values in the left
  panel are those of the models shown in fig.~\ref{fig-modelGrid_T},
  which are generated by fixing the mass value ($M=7.0M_\odot$) and
  varying the effective temperature. As evident from the figure, the
  best effective temperature is $T_{eff}=5320$ K.  The models plotted
  in the right panel correspond to those in
  fig.~\ref{fig-modelGrid_M}, obtained by fixing the effective
  temperature to the best fit value and changing the mass.
The inferred true distance modulus $\mu_0$, the $p$ factor and the
barycentric velocity $\gamma$ are labelled for each dot.

\begin{figure*}
\includegraphics[width=15.0cm]{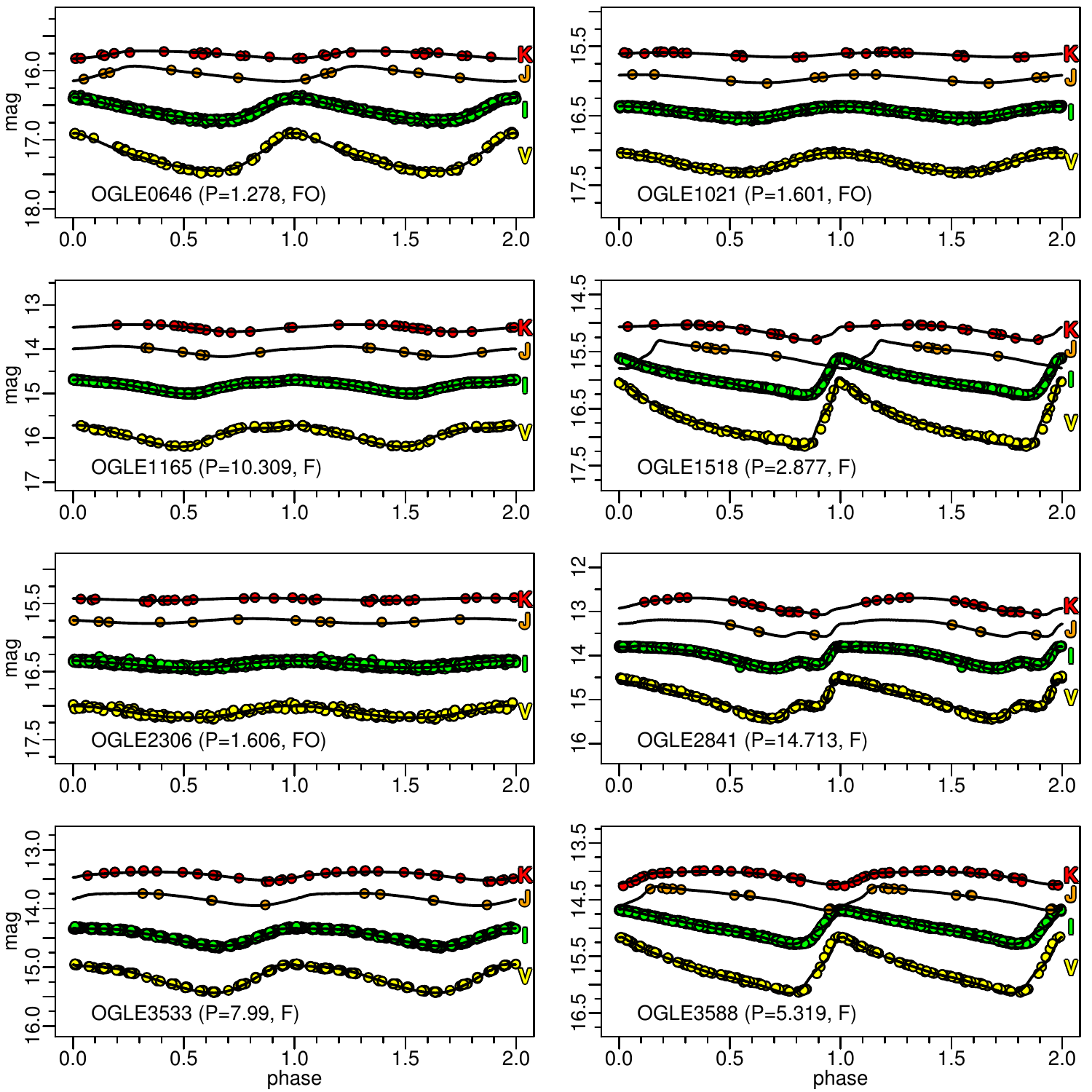}
\caption{Model fitting of selected SMC Cepheids for which only
  multi-filter light curves are available. The variable
  identification, pulsation period and mode are labelled in each
  panel. As the mean error bar is always $<0.02$ mag, we do not plot
  them for clarity reasons. The longest period Cepheid OGLE2470
    is shown in Fig.~\ref{fig-bestModels2} for clarity reasons.}
\label{fig-bestModels1}
\end{figure*}

\begin{figure*}
  \includegraphics[width=15.0cm]{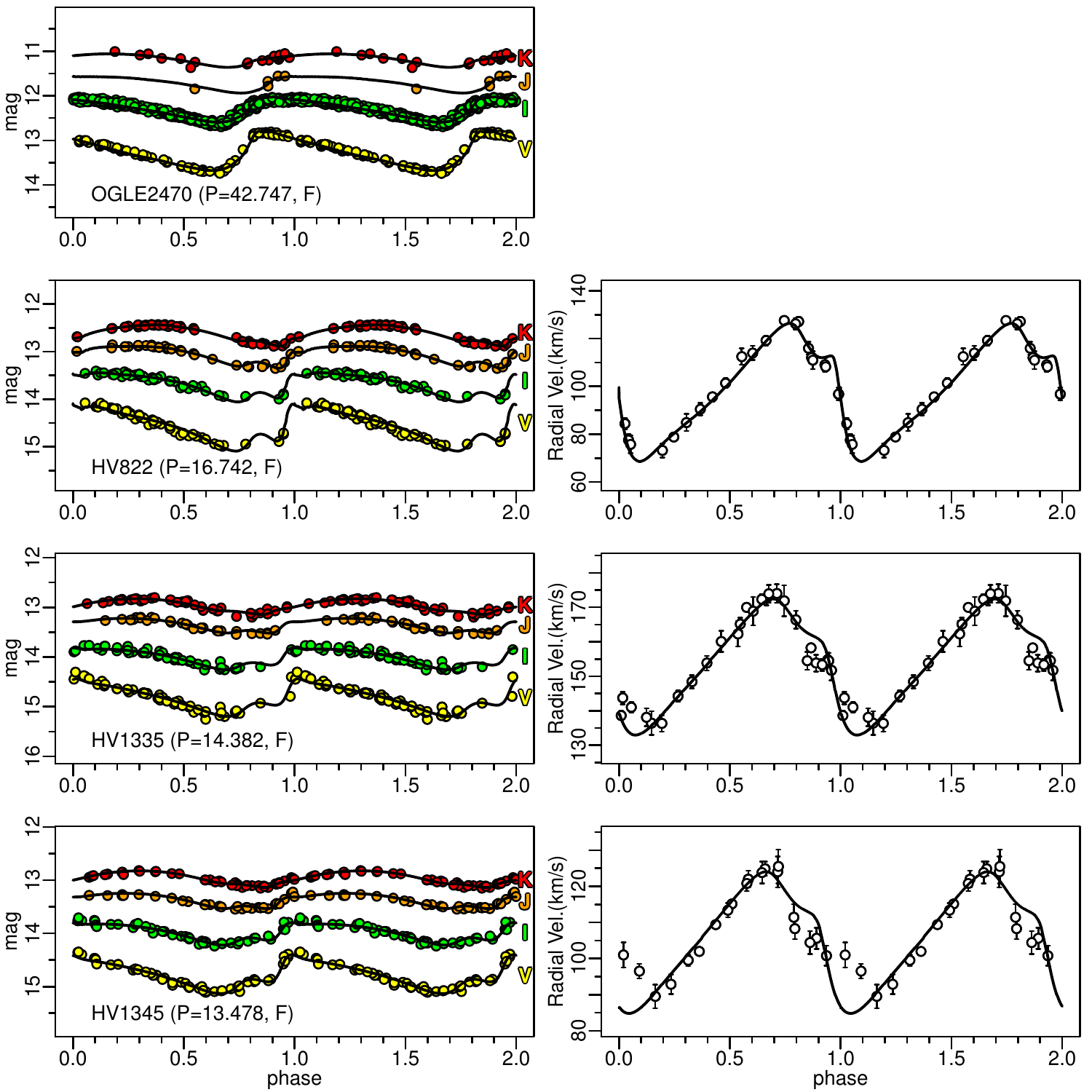}
\caption{Model fitting of  the photometric data of the longest period star OGLE2470 (left upper single panel) and  of the selected SMC Cepheids for which
  multi-filter light (left panels) and radial velocity (right panels)
  curves are available. The variable identification, pulsation period
  and mode are labelled in each panel. As the mean error bar for
  photometry  is always $<0.02$ mag we do not plot it for clarity reasons.}
\label{fig-bestModels2}
\end{figure*}

The same kind of analysis is applied to the other stars listed in
Table~\ref{tab:cepProperties} and the corresponding best fitting models are shown in Figures
\ref{fig-bestModels1} and \ref{fig-bestModels2}.

 We note that, at least for variables HV1335 and HV1345 the
 decreasing branch of the osberved radial
 velocity curve is not well reproduced by pulsation models. However,
 the good fit obtained for the $K$ band light curve, in a region of the
 spectrum where the radial variations dominate over the thermal
 effects, seems to suggest that the reason of the discrepancy is
 not to be ascribed to the pulsational computations. On the
   other hand, we cannot exclude the discrepancy  be due to the lack of description
   of the dynamical structure of the Cepheid atmosphere 
   (N. Nardetto, private communication; Nardetto et al. 2016 in press).
%%%The intrinsic stellar parameters of the obtained best fit models are
%%%reported in Table~\ref{tab:cepProperties}, where the errors are estimated according to the step in mass value and in effective temperature of the different sets of models \citep[see also][]{marconi13b}.
The intrinsic stellar parameters, namely the effective temperature,
the luminosity, the mass and the mean radius (see next Section), of the obtained best fitting models are
reported in columns 7-10 of Table~\ref{tab:cepProperties}. We give also an estimate of two different contributions to the error on the fitted parameters: one associated with the step in mass and temperature of the different sets of models, and the other due to the rms of observations around the best fitting models \citep[see also][]{marconi13b}.
The final 3 columns of the same Table report the inferred intrinsic distance modulus, p
factor and barycentric radial velocity with the associated incertainties. 
In the case of HV822 the inferred p factor is lower than the value
obtained for HV1335 and HV1345 in spite of the rather similar stellar
parameters. Indeed HV822 has almost the same effective temperature but
slightly higher mass and luminosity and lower gravity.  To test to
what extent this result depends on the photometric information, we also
tried to fit the radial velocity curve only and found a best fit model
with similar stellar parameters (fainter by 0.01 dex and less massive
by 0.1$M_{\odot}$), but a p factor around 1. On this basis we decided
to keep our originaly obtained best fit model, that correctly takes into account
all the available observations. A detailed investigation of the
predicted p factor dependence on stellar parameters is beyond the
purposes of the present paper and will be addressed in a future
publication (Molinaro et al. in preparation).

\section{Implications of results}

In this section we use the results obtained for the intrinsic stellar
parameters of the investigated CCs to determine constraints
both on the predicted MLR and PR relation as well as on the relations
that make CCs powerful standard candles, namely the PL and
Wesenheit relations, at least for the SMC chemical composition.

\subsection{The Mass-Luminosity relation}

\begin{figure}
\includegraphics[width=8.8cm]{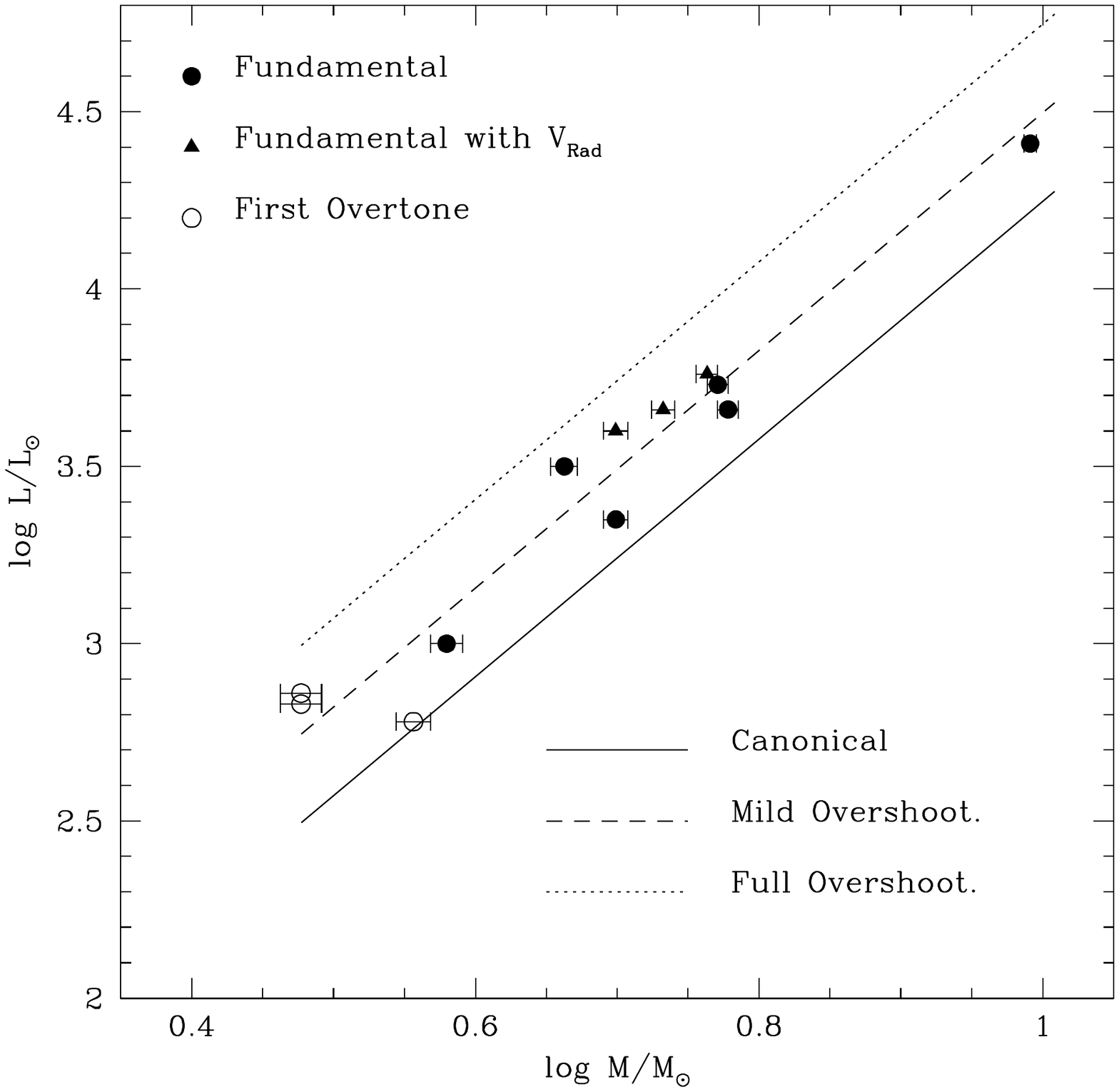}
\caption{Predicted MLR based on the model fitting results for
  both F (black solid circles) and FO Cepheids (open circles).  The
  best fitting model location in the MLR plane is compared with
   an evolutionary MLR obtained by neglecting both mass loss and core overshooting 
  and labelled as ``Canonical'' (solid line) and with the relations
  obtained by assuming  mild (dahed
  line; corresponding to an extension of the extra-mixing region beyond the Schwarzschild
border of about 0.2$H_p$ where $H_P$ is the pressure scale
height) or full  (dotted line) overshooting.}\label{ML}
\end{figure}

\begin{figure}
\includegraphics[width=8.8cm]{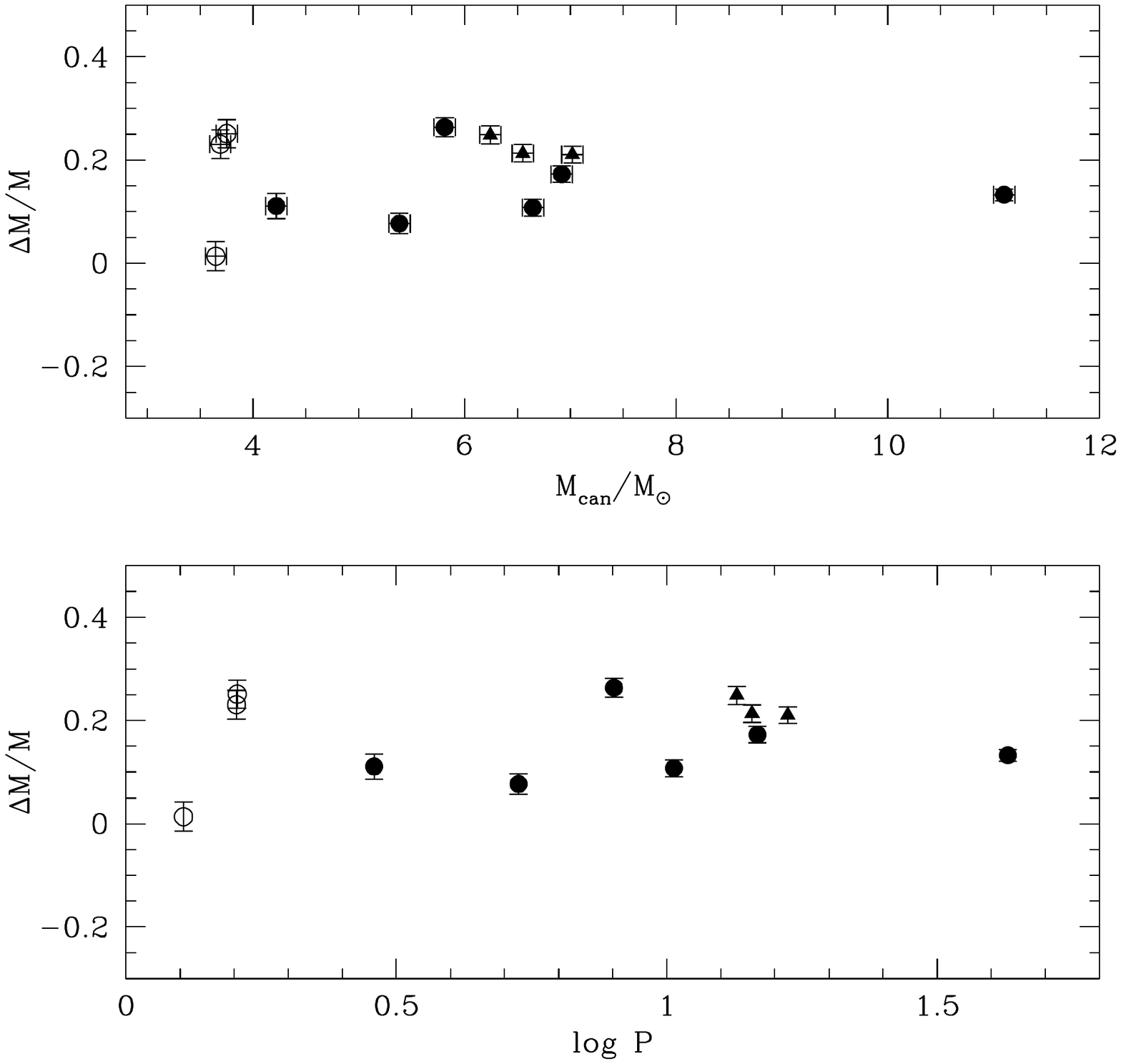}
\caption{Predicted deviation of the best fitting stellar
mass from the value corresponding to the canonical M for
  both F (black solid circles) and FO Cepheids (open circles).}\label{massloss}
\end{figure}

The stellar masses and luminosities of the best fitting models listed in
Table~\ref{tab:cepProperties} can be plotted in the MLR plane (see Figure \ref{ML}) and compared with current
predictions for the canonical and noncanonical MLRs.
The black solid and open symbols are the F and FO best fitting models
listed in Table~\ref{tab:cepProperties}, respectively. Their location in the MLR plane is
compared with the evolutionary predictions concerning the canonical
(no overshooting, no mass loss) MLR of \citet{bono00a} and
with the relations obtained by increasing the zero point of the
canonical one by 0.25 dex (dashed line) and 0.5 dex (dotted line) to
reproduce the effect of mild and full overshooting respectively
\citep[see e.g.][for details]{chiosi93,bms99}.
According to the points in Fig. \ref{ML} the investigated pulsators do not
follow a strict MLR, as expected in the case of canonical models or
constant overshooting effciency, but seem to favour a varying overluminosity with
respect to the canonical relation. Indeed, most of the Cepheids are
located between the mild and full overshooting lines.
Even if at this stage we cannot disentangle the role of overshooting
and mass loss in producing the quoted overluminosity, the detected
dispersion might indicate a combination of the two noncanonical
phenomena.
If overshooting were important, and this is a fundamental physics
aspect of stellar evolution, for a given mass, one would in principle
expect the same amount of overshooting (within small
uncertainties). The scatter we find in Fig. \ref{ML}, seems to imply
that another process, e.g. mass loss, should be important.  
On the other hand, if only mass loss were at work, this could be inferred from the
predicted deviation of the best fit stellar
mass from the value corresponding to the canonical MLR. Such a deviation
is represented in Figure \ref{massloss} as a function of the pulsation
period (bottom panel) and of the canonical mass (top panel) for the Cepheids in our sample. 
We note that the expected mass differences range from less than 2 \%
 to almost 30 \% and  are not clearly correlated neither
with the pulsation period nor with the stellar mass.
The dependence of  mass-loss efficiency on 
 the pulsation period is debated in the literature, with IRAS data
suggesting roughly constant values, and IUE spectra indicating a
dependence of the mass loss rate on the pulsation period but without
considering possible differences in the evolutionary times \citep[see
e.g.][and references therein]{caputo05,neils08,neils16}.
In conclusion, our results do not allow to disentangle between mass
loss and overshooting contributions to the observed overluminosity of
the investigated pulsators, leaving the possibility that a combination of the
two noncanonical phenomena might be at work.

\subsection{The Period-Radius relation}

\begin{figure}
\includegraphics[width=8.8cm]{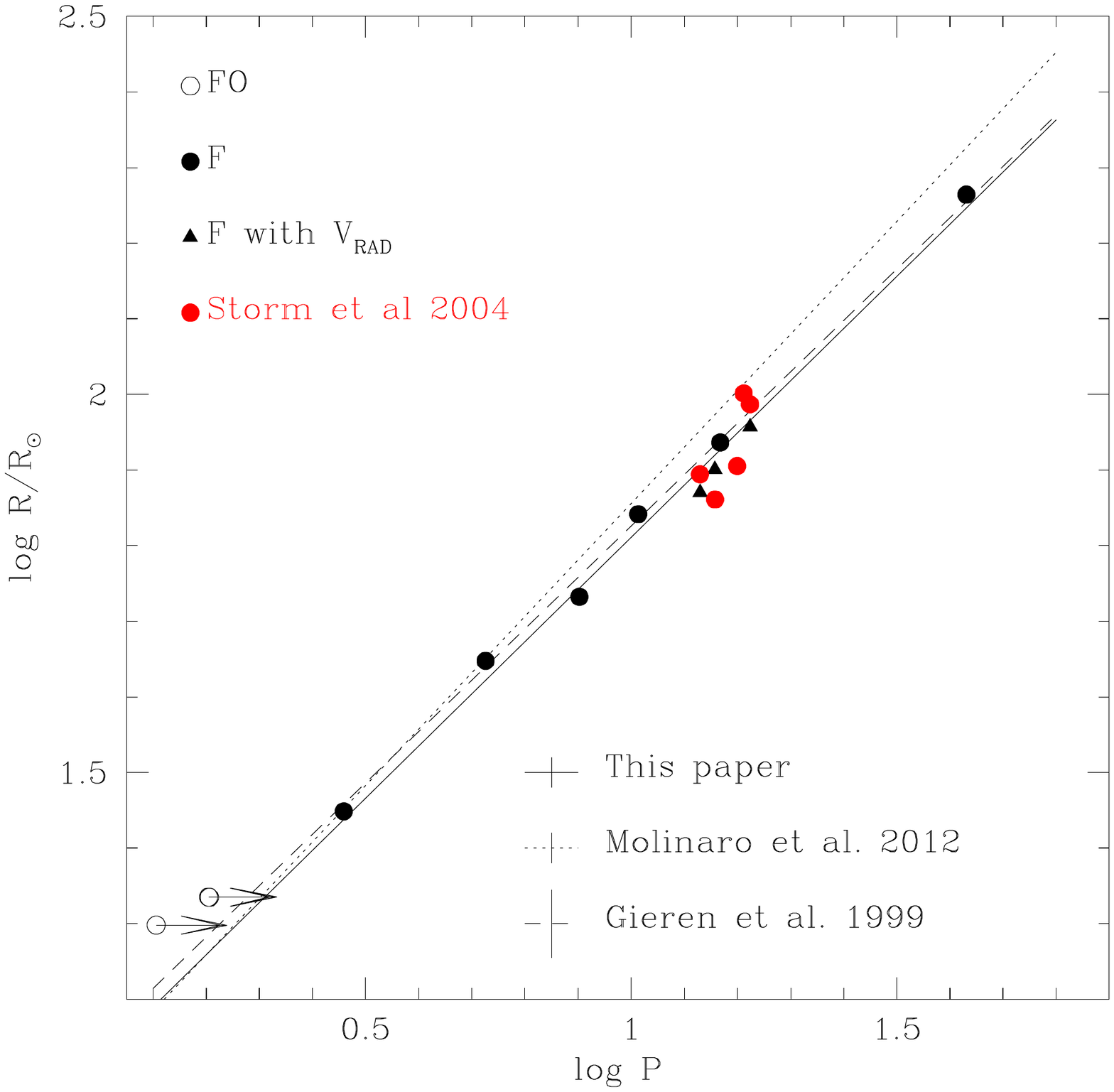}
\caption{Predicted PR relation based on the model fitting results for
  both F (black solid circles) and FO Cepheids (open circles). The
  latter have been fundamentalized (see arrows) before performing the
  linear regression (solid line). The obtained PR has been compared
  with the location of SMC  Cepheids with radius estimates by
  \citet{storm04} and with the relations by \citet[][dotted line]{molinaro12} and \citet[][dashed line]{gieren99}.}
\label{PR}
\end{figure}

The adopted nonlinear hydrodynamical code also allows us to model the
variation of radius along the pulsation cycle for each pulsation
model. Once the radius curve is obtained we are able to derive the
time averaged mean radius and to correlate it with the corresponding
pulsation period. The location of the eleven best fitting models obtained
in the previous Section in the Period-Mean Radius diagram is shown in
Fig. \ref{PR}.

In this plot F and FO best fit models are represented by black solid
and open circles respectively; the red symbols correspond to the
SMC observed classical Cepheids with radii estimated by \citet{storm04}.
The solid line is the theoretical linear regression for the complete
sample where the three FO periods  have been
fundamentalized\footnote{We considered the period the star would have if it were
  a F pulsator. This is computed by the linear pulsation code that is
  adopted to evaluate the radial eigenfunctions and to provide the
  envelope structure for subsequent nonlinear pulsation
  computations\citep[see][and references therein]{bms99}}.
Unfortunately, no empirical SMC PR relation has been published in the
literature, to be compared with our theoretical relation.
The dotted  line depicts the linear relatios derived by
\citet{molinaro12} on the basis of the CORS Baade-Wesselink method
applied to a sample of 11 Cepheids belonging to the young LMC blue
populous cluster NGC 1866, while the dashed line is the relation derived by \citet{gieren99}
 for a sample of both Galactic and Magellanic Cepheids.
We note that the predicted PR relation based on the model fitting of SMC
Cepheids suggests that at fixed period the radius is smaller for SMC
Cepheids than for Galactic and LMC ones. This result was already
obtained  by  \citet{storm04} on the basis of the comparison
presented in that paper between the radii of SMC and Galactic
Cepheids.

\subsection{The PL relations}

\begin{figure}
\includegraphics[width=8.8cm]{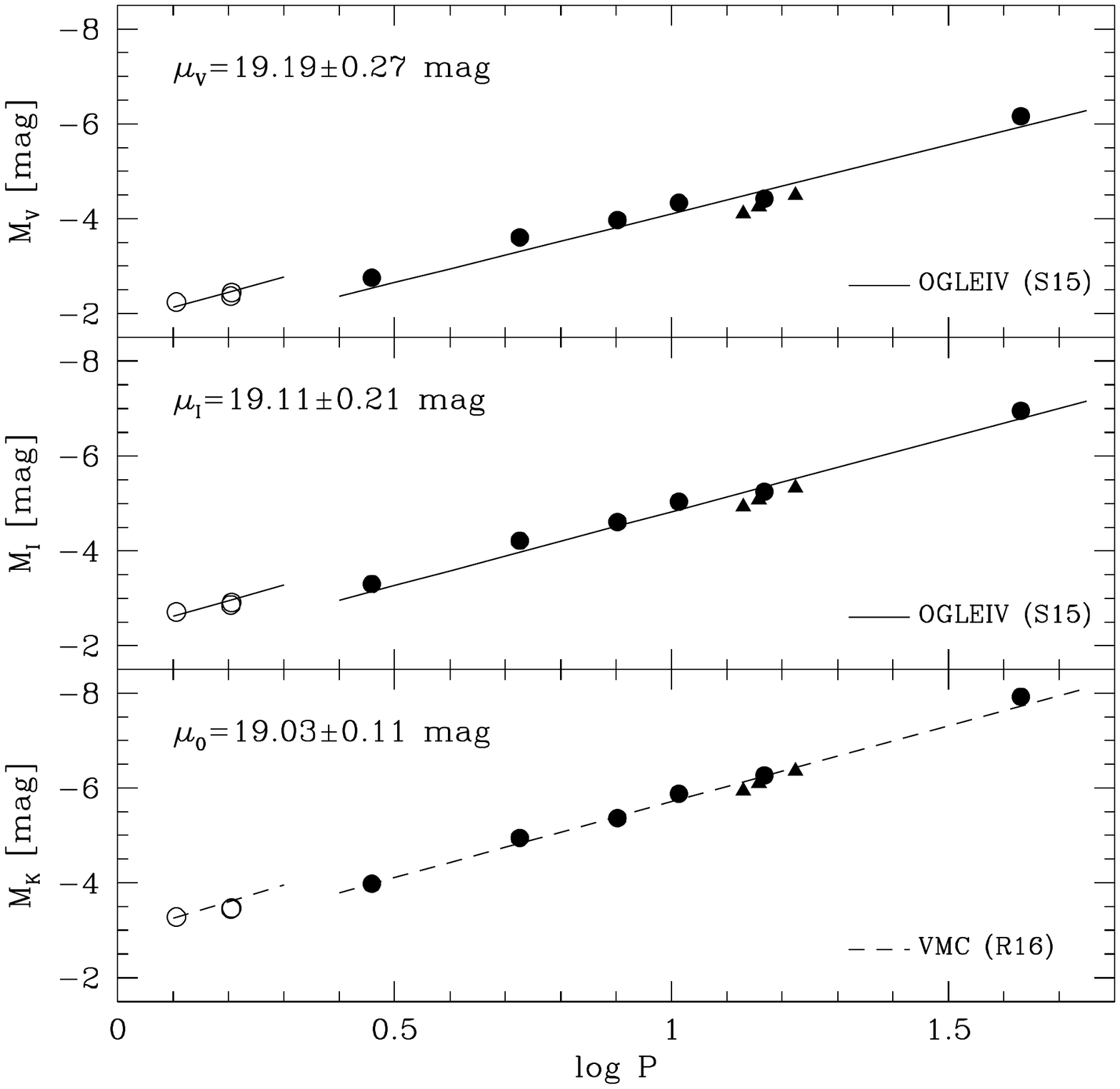}
\caption{Predicted PL relation in the
  V (top panel), I (middle panel) and $K$ (bottom panel) bands based on the model fitting results for
  both F (black solid circles) and FO (open circles) Cepheids. The
  empirical relations by the OGLE IV (solid lines) and
  VMC (dashed line) collaborations are shown for comparison and used
  in combination to the predicted absolute magnitudes to infer the
  labelled distance moduli.
}
\label{PL}
\end{figure}

The mean magnitudes of the best fitting models can be correlated
with the corresponding periods to build multifilter PL
relations.
In Fig. \ref{PL} we show the location of both F (red solid circles) and FO
(blue open circles)  best fitting models in the
V, I and $K$ band Magnitude versus Period planes. The PL relations
recently obtained by the OGLE IV collaboration \citep[][S15 in the
label]{sosz15} and by the VMC collaboration  \citep[][R16 in the
label]{ripe16} for larger Cepheid samples including our targets, are overplotted for the average distance moduli
obtained in the various bands by subtracting these relations to the Cepheid absolute
magnitudes. As the empirical VMC relations have been corrected for
reddening \citep[see][for details]{ripe16} their application to the absolute
$K$ band magnitudes of our sample directly provides the instrinsic
distance modulus $\mu_0$ (see label).

On the other hand the difference between the obtained $\mu_V$ and
  $\mu_I$  provides and independent estimate of the average color excess $E(V −I) = 0.14 \pm 0.12$ mag, where the error is the standard
  deviation. This value is in agreement, within the errors, with the
  literature estimates
  \citep[see e.g.][and references therein]{HGD11,more16,ripe16}. 
We also note that the theoretical slopes agree quite well with the
empirical ones, for both F and FO pulsators.  Moreover,
 the inferred intrinsic mean distance modulus is equal to $\mu_0=19.01$ mag and its standard deviation to 0.08 mag, in agreement with the most
recent literature values \citep[see][and references
therein]{degr15,ripe16}.

\subsection{The Wesenheit relations}

Finally, it is interesting to compare the predicted properties of the
SMC Cepheids in our sample with currently adopted Wesenheit
relations. 
These are reddening-free Period-Luminosity-Color relations, obtained by
fixing the colour term coefficient to the ratio between total to
selective extinction in the considered filters \citep[see e.g.][and
references therein]{madore82,cmm00,bono10,fmm13,ripe16}.
In particular, to exploit both the optical and the NIR data, we adopt
$W(V,I)=I-1.55\times(V-I)$ and $W(V,K)=K-0.13\times(V-K)$ according to recent 
prescriptions in the literature \citep[see e.g.][and references therein]{sosz15,ripe16}.

\begin{figure}
\includegraphics[width=8.8cm]{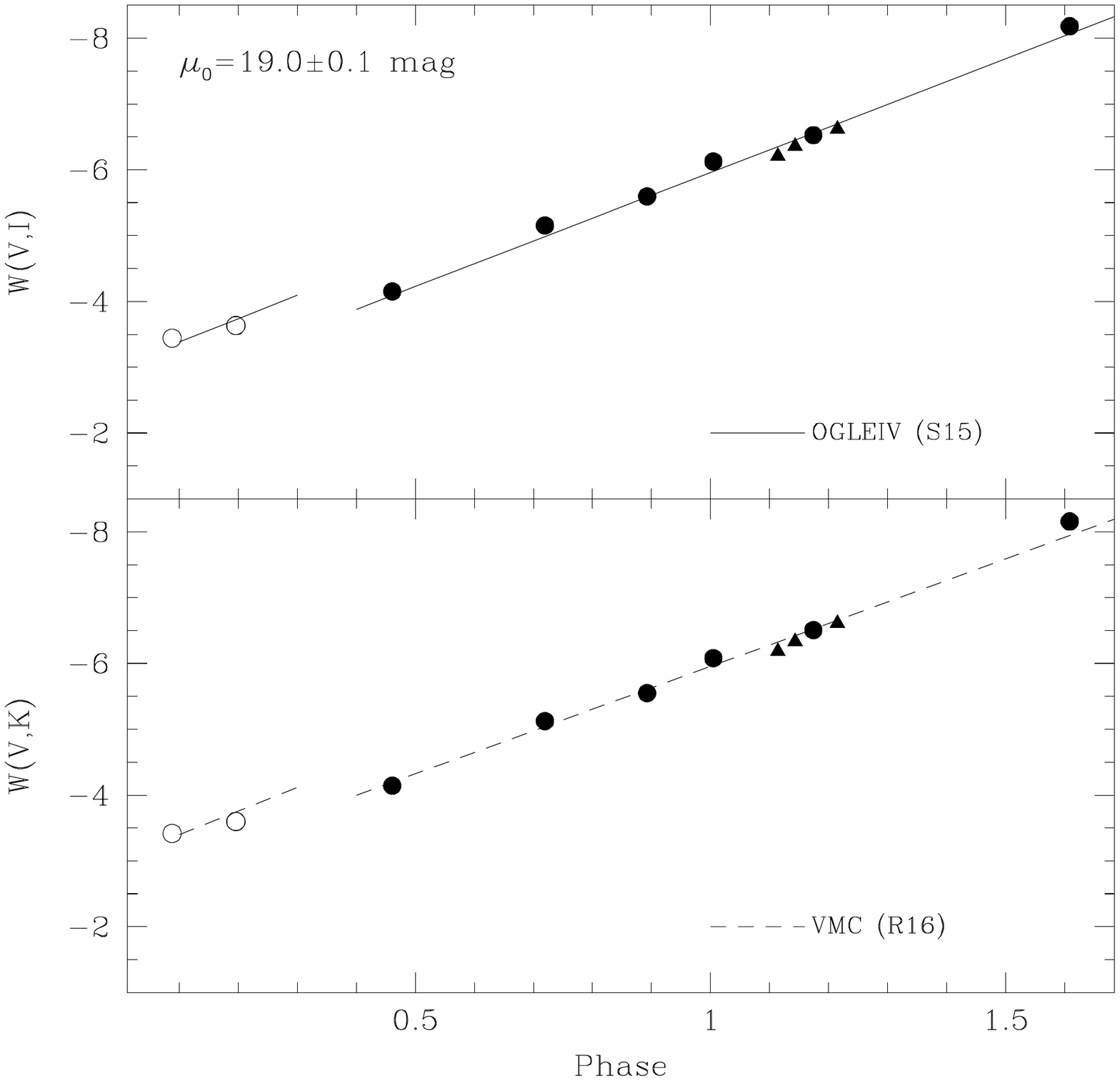}
\caption{Predicted PW relations in the "$V$, $I$" (top panel) and "$V$, $K$" (bottom
  panel) filters for both F and FO pulsators, compared with the
  empirical relations by \citet{sosz15} and
\citet{ripe16}.}\label{pw}
\end{figure}

As shown in Fig.~\ref{pw} the models that best reproduce the light curves of
the selected SMC Cepheids nicely agree with the empirical SMC
Wesenheit  relations recently presented by  \citet{sosz15} and
\citet{ripe16}, for an inferred distance modulus ($\mu_0=19.0\pm0.1$ mag)  in excellent agreement with the values quoted above.

\section{Summary and Future Perspectives}\label{sec-conclusions}

We have presented the multiwavelength optical and NIR light curve model fitting for a
sample of 12 Cepheids (9 fundamental and 3 first overtone pulsators)
in the SMC, in order to constrain their instrinsic stellar parameters
and distances. The optical photometry was taken from data of the OGLE
collaboration, while the near-infrared photometry ($JK$ filters) is
based on VMC observations.
For three stars radial velocity curves were also available and the
best fitting model was obtained by combining photometric and radial
velocity data.

Following this approach, we were able to derive the intrinsic
properties and the individual distances for the Cepheids in our
sample, and in turn to find that:
\begin{itemize}

\item The instrinsic masses and luminosities of the  inferred best fitting models
show that all these pulsators are brighter than the canonical
evolutionary MLR suggesting a significant
efficiency of core overshooting and/or mass loss. Assuming that
the inferred deviation from the canonical MLR is only due to
mass loss, we also discussed the distribution of the percentage mass loss as a
function of the pulsation period and the canonical stellar mass.

\item The inferred individual distances provide a mean value for the
  SMC distance modulus in agreement with the literature and a
  dispersion that can be ascribed to real variations in distance
  within the SMC. Indeed, there is evidence in the literature that the
  depth of the SMC is up to $\sim$ 0.3 mag \citep[see e.g.][and
  references therein]{glatt08,sub09}.

\item The obtained stellar radii can be correlated with the periods to built
a PR relation that is found to be in excellent agreement with current
PR relations derived in the literature for SMC Cepheids.

\item The absolute magnitudes of the best fitting models were combined with the
period information to show the behaviour of the investigated stars in
the PL and Period-Wesenheit planes, finding an excellent agreement with
published relations based on VMC data.

\end{itemize}

The results of this investigation support the predictive capabilities
of the adopted theoretical scenario in terms of individual distances
and intrinsic stellar parameters and pave the way to the
application to other extensive databases at various chemical
compositions, including the VMC Large Magellanic Cloud pulsators (Ragosta et
al. in preparation).

In the future we also plan to extend the application to first overtone
pulsators in order to better constrain their PL and  Period-Wesenheit
relations and to test the accuracy of the method through
application to the light curves of Galactic CCs with Gaia
parallaxes \citep{gaia1,gaia2}. The latter comparison, once fixed the distance to the Gaia
results,  will also allow us to put strong constraints
on the physical and numerical assumptions adopted in the
hydrodynamical code as well as on the predicted stellar masses, MLR,
and, provided that the  metallicity is precisely constrained by complementary spectroscopic data,
on the helium to metal enrichment ratio.

\section*{Acknowledgements}
We thank our referee, N. Nardetto, for his useful comments and
  suggestions. We also deeply thank J. P.  Emerson for a critical reading of the paper and
valuable comments.
This work was supported by PRIN INAF 2014 (P.I.: G. Clementini). We thank the Cambridge Astronomy Survey Unit (CASU) and the Wide Field Astronomy Unit (WFAU) in Edinburgh for providing calibrated data products under the support of the Science and Technology Facility Council (STFC) in the UK.
M.R.C acknowledges support from the UK's
Science and Technology Facilities Council (grant number
ST/M00108/1).
RdG acknowledges research support from the National Natural Science
Foundation of China (NSFC) through grants 11373010, 11633005 and
U1631102. This project has received funding from the European Research Council (ERC) under the European Union's Horizon 2020 research and innovation programme (grant agreement No 682115).

%%%%%%%%%%%%%%%%%%%%%%%%%%%%%%%%%%%%%%%%%%%%%%%%%%



%\bibliographystyle{mnras}


%%%%%%%%%%%%%%%%%%%%%%%%%%%%%%%%%%%%%%%%%%%%%%%%%%

%%%%%%%%%%%%%%%%% APPENDICES %%%%%%%%%%%%%%%%%%%%%

%\appendix

% Don't change these lines
\bsp	% typesetting comment
\label{lastpage}
\end{document}